\documentclass[
 reprint,
 superscriptaddress,
 preprintnumbers,
 amsmath,
 amssymb,
 prl,
 floatfix,
 aps,
]{revtex4-2}

\usepackage[utf8]{inputenc}
\usepackage[final]{changes}
\usepackage[english]{babel}
\usepackage[T1]{fontenc}
\usepackage{amsfonts}
\usepackage{amssymb}
\usepackage{graphicx}
\usepackage{braket}
\usepackage{verbatim}
\usepackage{hyperref}
\usepackage{subfigure}
\usepackage{longtable}
\usepackage{tabularx}
\usepackage{dcolumn}%
\usepackage{color}
\usepackage{graphicx}% Include figure files
\usepackage{dcolumn}% Align table columns on decimal point
\usepackage{bm}% bold math
\usepackage{comment}
\usepackage{booktabs}
\usepackage{makecell}
\usepackage{siunitx}
\usepackage{xcolor}
\newcommand{\bs}[1]{\boldsymbol{#1}}
\newcommand{\dd}{\mathrm{d}} 

\begin{document} 

\author{Stefano Paolo Villani} 
\email[]{svillani@uchicago.edu} 
\affiliation{Pritzker School of Molecular Engineering, University of Chicago, Chicago, Illinois 60637, USA} 

\author{Yu Jin} 
\affiliation{Pritzker School of Molecular Engineering, University of Chicago, Chicago, Illinois 60637, USA}
\affiliation{Initiative for Computational Catalysis, Flatiron Institute, New York, NY 10010, USA} 

\author{Giulia Galli} 
\email[]{gagalli@uchicago.edu} 
\affiliation{Pritzker School of Molecular Engineering, University of Chicago, Chicago, Illinois 60637, USA} 
\affiliation{Department of Chemistry, University of Chicago, Chicago, Illinois 60637, USA} 
\affiliation{Materials Science Division and Center for Molecular Engineering, Argonne National Laboratory, Lemont, Illinois 60439, USA} 

\title{First-principles calculations of internal conversion processes in spin defects} 

\begin{abstract} 
Optically active spin defects are foundational for quantum technologies, yet common approximations underestimate their internal conversion (IC) rates by orders of magnitude. We propose a broad, predictive framework to compute IC rates that incorporates multi-configurational effects via many-body wavefunctions in TDDFT, and includes all-phonon-mode contributions via analytical non-adiabatic couplings. Our approach resolves discrepancies with experiment, achieving quantitative agreement for the NV$^-$ center in diamond, and identifying a previously overlooked non-radiative channel in the divacancy triplet lifetime in SiC.
\end{abstract} 

\maketitle 

\begin{figure}[!h]
    \begin{minipage}[c]{1.0\linewidth}
    \centering
    \includegraphics[width=.85\textwidth]{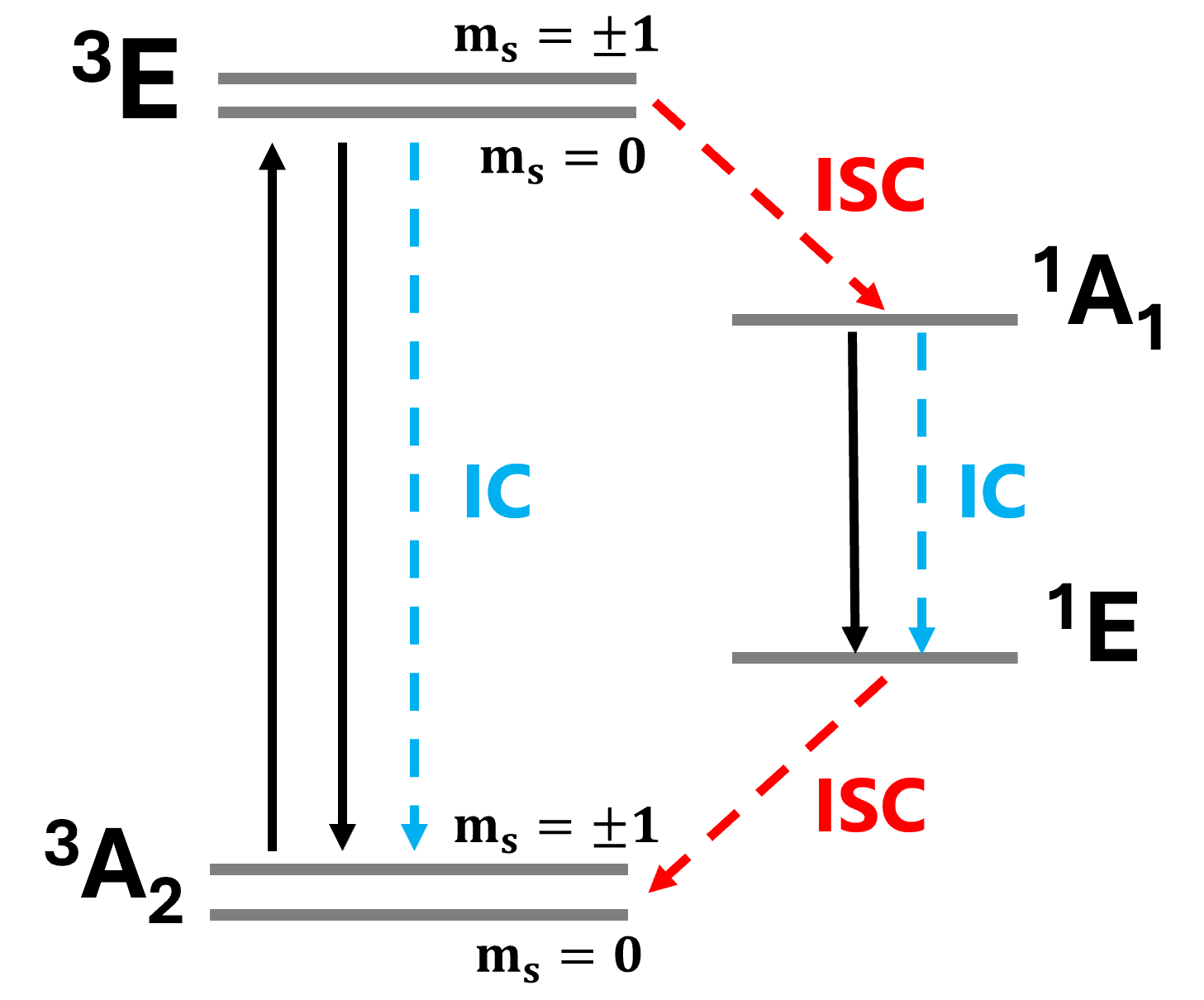}
    \end{minipage}
    \caption{Schematic representation of the electronic energy levels of the NV${^-}$ center in diamond and of the VV$^0$ center in silicon carbide. Both systems exhibit a triplet ground state ($^3$A$_2$) and triplet excited state ($^3$E), and two singlet shelving states, $^1$A$_1$ and $^1$E, all labeled according to the C$_{3v}$ point group. In an optical cycle, after a laser excitation from $^3$A$_2$ to $^3$E, the system can either return to the ground state or transition to $^1$A$_1$ through  inter-system crossing (ISC); from $^1$A$_1$ the electron then relaxes to $^1$E, through non-radiative spin-conserving transitions (internal conversion, IC), and finally to the ground state through a second ISC process. The magnetic sublevels of the ground state represent an effective qubit system\cite{PhysRevB.98.085207}. Radiative and non-radiative transitions are indicated by solid and dashed lines, respectively.}
    \label{fig:optical_cycle}
\end{figure}

\textit{Introduction}---Understanding and manipulating the properties of defects in solids has driven technological innovation for decades. Recently, optically active spin defects, e.g., the negatively charged nitrogen-vacancy (NV$^-$) center in diamond
% \cite{walker1979optical,doherty2013nitrogen} 
and the neutral divacancy (VV$^0$) center in silicon carbide (SiC)
% \cite{son2020developing,falk2013polytype}, 
have emerged as promising platforms for the development of quantum technologies\cite{
% weber2010quantum,
chatterjee2021semiconductor,wolfowicz2021quantum,anderson2022five,fang2024quantum,nishikawa2025coherent,li2025non,roberts2025quantum,tesiman2026surveying}.
% are effective spin qubits, and they hold promises for applications in quantum computing\cite{weber2010quantum,waldherr2014quantum}, sensing\cite{schirhagl2014nitrogen,barry2020sensitivity}, and  communication\cite{childress2013diamond,christle2017isolated,wolfowicz2017optical,anderson2022five}. 
\\ \indent 
Both experiments and first-principles, theoretical studies have shed light on many aspects of the NV$^-$ and VV$^0$ optical cycle\cite{PhysRevB.98.085207,thiering2017ab,bian2025theory,alkauskas2014first_NV,jin2021photoluminescence,jin2022vibrationally,jin2025first,turiansky2026frontiers,li2026computation}, illustrated in Figure \ref{fig:optical_cycle}. Radiative transitions and inter-system crossing rates have been investigated using various first-principles frameworks\cite{alkauskas2014first_NV,jin2021photoluminescence,PhysRevB.98.085207,jin2022vibrationally,razinkovas2021vibrational,jin2025first,squires2026guidelines,galli2026strategies}; however, internal conversion (IC) processes remain much less explored, thus hampering a full description of the optical cycle from first principles. These non-radiative processes, arising from thermal excitations due to the coupling of electronic and nuclear degrees of freedom, are key to understanding the effect of temperature on the optical cycle\cite{bian2025theory} and to predict excited-state lifetimes\cite{robledo2011spin,choi2012mechanism,bian2025theory}. A detailed description of the lifetimes of excited states, and of the competition between radiative and non-radiative processes, is necessary to determine the efficiency of the readout process\cite{ping2021computational} in an optical cycle, and to interpret and predict optically-detected magnetic resonance (ODMR) spectra\cite{li2024excited}. 
\\ \indent
The central challenges in describing IC processes are the accurate description of the many-body excited states of the defects and of their electron-phonon (e--ph) coupling. \textit{Ab-initio} methods based on one-dimensional effective modes and Kohn-Sham electronic orbitals have been successfully used in the literature to describe non-radiative capture rates of charged carriers in several semiconductors and insulators in the strong-coupling regime\cite{alkauskas2014first_NR,turiansky2021nonrad,shi2015comparative,alkauskas2016tutorial,wickramaratne2018comment}. 
% non-radiative capture processes of charge carriers in defective semiconductors and insulators
However, their application to neutral excitations in spin-defective solids resulted in inaccurate lifetimes' estimates\cite{li2024excited,kazemi2026giant}
% presumably due to an inadequate description of both the many-body nature of the electronic states of the defects and of multi-phonons processes.
% % The complete understanding of ICs in spin-defects from first principles has thus been hampered by the theoretical complexity and computational burden of obtaining e--ph couplings between many-body electronic states and all phonon modes. 
% Evidence: Ping et al underestimate the IC rate of the singlets transition in NV-; recent results on the T-center in Silicon regarding the use of effective mode approximation. Finally, The seminal paper by Van-de-Walle urgently calls for methods to go beyond effective-mode approximations.
\\ \indent
In this Letter, we present an efficient and accurate \textit{ab-initio} framework for the calculation of non-radiative rates of internal conversion processes in spin defects. We evaluate the non-adiabatic coupling (NAC) between electronic excited states using linear-response time-dependent density functional theory (LR-TDDFT) with hybrid functionals to capture the multi-configurational character of the excited states, missing in single-orbital descriptions. In addition, we compute analytical derivatives of non-adiabatic couplings, thus enabling the inclusion of all phonon modes in the calculations of rates. We apply the framework to study IC processes in the optical cycles of the NV$^-$ center in diamond and the VV$^0$ center in SiC. For the first time, we obtain results in excellent agreement with experiments for the NV$^-$ and we identify a substantial non-radiative contribution to the decay of the triplet excited state of the VV$^0$ center, resolving a long-standing discrepancy with experiments.

\textit{Theoretical framework}---We describe the internal conversion processes as non-radiative transitions between two many-body electronic states with the same spin, occurring via multi-phonon exchange\cite{huang1950theory,kubo1955application,passler1975calculation,henry1977nonradiative,abakumov1991nonradiative,imbush1991advances}.
We define a temperature-dependent non-radiative transition rate (NRTR) $\Gamma_{\mathrm{NR},IF}(T)$ between an initial ($I$) and a final ($F$) state, using the Fermi's golden rule at first-order in the perturbing e--ph interaction\cite{stoneham1975theory,kubo1955application,wu2019carrier,nichols2025multiphonon}: 
\begin{equation}
    \label{eq:def_NRTR}
    \Gamma_{\mathrm{NR},IF}(T) = \frac{2\pi}{\hbar}\sum_k^{N_\mathrm{modes}} |M_{IF}^k|^2 X_{IF}^k(T).
\end{equation}
Eq. (\ref{eq:def_NRTR}) accounts for the contribution of all the $N_\mathrm{modes}$ phonon modes of the solid. The term $M_{IF}^k$ defines the e--ph matrix element between the initial and final electronic states and measures the strength of the coupling with mode $k$. The term $X_{IF}^k(T)$ is the phonon-relaxation contribution of the $k$-th mode due to the rearrangement of the atoms from the relaxed geometry $\bs{R}_I$ of state $I$ to that of state $F$ ($\bs{R}_F$); this term reads:
\begin{equation}
    \label{eq:def_Xif}
    X_{IF}^k(T) = \sum_{m,n} \rho_n(T) | \langle \Phi_{I,n} | Q_k | \Phi_{F,m} \rangle |^2 \times \delta(E_{I,n}^{\mathrm{tot}}  - E_{F,m}^{\mathrm{tot}}) \,,
\end{equation}
where $|\Phi_{I,n}\rangle$ ($|\Phi_{F,m}\rangle$) is the multi-phonon wave-function of the vibrational configuration $n$ ($m$) of the electronic state $I$ ($F$). 
% Indeed, the IC process involves multi-phonon exchange between an initial $n$
% =\{n_1,\dots,n_{N_\mathrm{modes}}\}$ 
% and a final $m$ vibrational states.
% =\{m_1,\dots,m_{N_\mathrm{modes}}\}$, with $n_j$ and $m_j$ being the occupation numbers of mode $j$. 
The initial state $n$ is assumed to be the equilibrium distribution $\rho_n(T)$ at temperature $T$, and we consider all possible final vibrational states compatible with the conservation of the total energy, nuclear plus electronic. This requirement is enforced through the delta-function in Eq. (\ref{eq:def_Xif}), where:
% \begin{align}
% %     E_{I,n}^{\mathrm{tot}} &= E_I(\bs{R}_I) + \sum_j^{N_\mathrm{modes}}\left(n_j+\frac{1}{2}\right)\hbar\omega_j \,, \\
% %     E_{F,m}^{\mathrm{tot}} &= E_F(\bs{R}_F) + \sum_{j}^{N_\mathrm{modes}}\left(m_{j}+\frac{1}{2}\right)\hbar\omega'_{j}.
%     E_{L,l}^{\mathrm{tot}} = E_L(\bs{R}_L) + \sum_j^{N_\mathrm{modes}}\left(l_j+\frac{1}{2}\right)\hbar\omega_{j}^L\,,
% \end{align}
% with $L,l=I,n$ or $L,l=F,m$.
\begin{equation}
\begin{aligned}
    E_{I,n}^{\mathrm{tot}}  - E_{F,m}^{\mathrm{tot}} = \Delta_{IF}
    &+ \sum_j^{N_\mathrm{modes}}\left(n_j+\frac{1}{2}\right)\hbar\omega_{j}^I \\ &- \sum_j^{N_\mathrm{modes}}\left(m_j+\frac{1}{2}\right)\hbar\omega_{j}^F \,,
\end{aligned}
\end{equation}
with $\Delta_{IF}=E_I(\bs{R}_I) - E_F(\bs{R}_F)$ the adiabatic energy difference. Hereafter, we adopt the displaced harmonic oscillator approximation and consider $\hbar\omega^F_j=\hbar\omega^I_{j}$.
\\ \indent
The e--ph matrix element $M_{IF}^k$ in Eq. (\ref{eq:def_NRTR}) is defined as
\begin{equation}
    \label{eq:def_Mif}
    M_{IF}^k = \langle \Psi_I| \frac{\partial H_\mathrm{el} }{\partial Q_k} | \Psi_F \rangle \,,
\end{equation}
where $|\Psi_I\rangle$ and $|\Psi_F\rangle$ are the many-body wavefunctions of the initial and final electronic states, respectively, $H_\mathrm{el}$ is the electronic many-body Hamiltonian, and $Q_k$ is the normal coordinate of mode $k$. All quantities are computed at $\bs{R}_I$, using the following relations:
\begin{align}
    \frac{\partial H_\mathrm{el}}{\partial Q_k} &= \sum_{A}^{N_\mathrm{atoms}}\nabla_{\bs{R}_A} H_\mathrm{el}\cdot \frac{\bs{\mathrm{e}}^k_{A}}{\sqrt{M_A}} \,, \\
    \label{eq:eph_to_nac}
    \langle \Psi_I|\nabla_{\bs{R}_{A}} H_\mathrm{el} | \Psi_F\rangle  &= (E_F - E_I)\langle \Psi_I|\nabla_{\bs{R}_{A}} \Psi_F\rangle \,,
\end{align}
where $M_A$ is the atomic mass, $\bs{\mathrm{e}}^k$ is the $k$-th eigenvector of the dynamical matrix, and $E_I$ and $E_F$ are the energies of the initial and final electronic states, respectively. We rewrite Eq. (\ref{eq:def_Mif}) as 
\begin{equation}
    \label{eq:def_Mif_NACs}
    M_{IF}^k = (E_F - E_I) \sum_A^{N_\mathrm{atoms}} \bs{d}_{IF,A} \cdot \frac{\bs{\mathrm{e}}^k_{A}}{\sqrt{M_A}} \,,
\end{equation}
where we have introduced the non-adiabatic coupling (NAC) vectors $\bs{d}_{IF,A}$, defined as 
\begin{equation}
    \label{eq:def_NACs}
    \bs{d}_{IF,A} = \langle \Psi_I|\nabla_{\bs{R}_{A}} \Psi_F\rangle.
\end{equation}
The NACs are central quantities in the description of ultra-fast photo-chemical and photo-physical processes, especially when two adiabatic potential energy surfaces are close to each other or cross\cite{herzberg1963intersection,yarkony2001conical,matsika2011nonadiabatic,matsika2021electronic}. They also play a key role in non-adiabatic molecular dynamics\cite{tully1990molecular,baer2006beyond,tapavicza2007trajectory,curchod2013trajectory}. Here, we use Eq. (\ref{eq:def_Mif_NACs}) to compute the e--ph matrix elements in terms of the vectors $\bs{d}_{IF,A}$. We implemented the calculation of NACs in the LR-TDDFT framework in the WEST code\cite{govoni2015large,jin2022vibrationally,jin2023excited} using an analytical formulation based on the extended Lagrangian approach\cite{li2014first_theory,li2014first_implementation,wang2021nac,liu2023exciton,villani2026}. Hence, we avoid the use of finite differences to compute wave-function derivatives, and we can efficiently include all phonon modes while accounting for the many-body nature of the electronic states. Our implementation is optimized for GPU platforms and supports hybrid functionals and spin-flip excitations\cite{yu2022gpu,jin2022vibrationally,jin2023excited,yu2025many,villani2026}.
\\ \indent
To compute the NRTRs from Eq. (\ref{eq:def_NRTR}), we adopt the generating function approach \cite{kubo1955application,nichols2025multiphonon}. We define a line-shape function $F_{\mathrm{NR}}(\Delta)$ (the subscripts $I$ and $F$ and the temperature $T$ are omitted here):
\begin{align}
    % F_{\mathrm{NR}}(\Delta) 
    % &= \frac{2\pi g}{\hbar} \Re \int_{-\infty}^{\infty} D(E') G(\Delta - E') \dd E',  
    F_{\mathrm{NR}}(\Delta) 
    &= \frac{g}{\hbar^2}\sum_k^{N_\mathrm{modes}} |M_{IF}^{k}|^2  \Re \int_{-\infty}^{\infty}  B_k(E') G(\Delta - E') \dd E', 
    % \int_{-\infty}^{\infty} B_k(t) e^{\frac{iE't}{\hbar}} \dd t
\end{align}
where $g$ is the degeneracy of the final state\cite{alkauskas2014first_NR}, and $B_k(E)$ and $G(E)$ are $T$-dependent spectral functions 
% , previously adopted to describe radiative transitions\cite{jin2021photoluminescence,jin2022vibrationally}
% \begin{equation}
%     \label{eq:def_D(E)}
%     D(E) = \frac{1}{2\pi\hbar} \sum_k^{N_\mathrm{modes}} |M_{IF}^{k}|^2 \int_{-\infty}^{\infty} B_k(t) e^{\frac{iEt}{\hbar}} dt 
% \end{equation}
% is the spectral density for the electron-phonon coupling, 
% and $B_k(E')$ is $T$-dependent and 
that accounts for multi-phonon processes (See Appendix). Finally, we obtain the value of the NRTR of a given IC process when the energy $\Delta$ equals the adiabatic energy difference $\Delta_{IF}$, namely $\Gamma_{\mathrm{NR},IF}=F_{\mathrm{NR}}(\Delta_{IF})$.

\begin{figure}[!h]
    \begin{minipage}[c]{1.0\linewidth}
    \centering
    \includegraphics[width=1.0\textwidth]{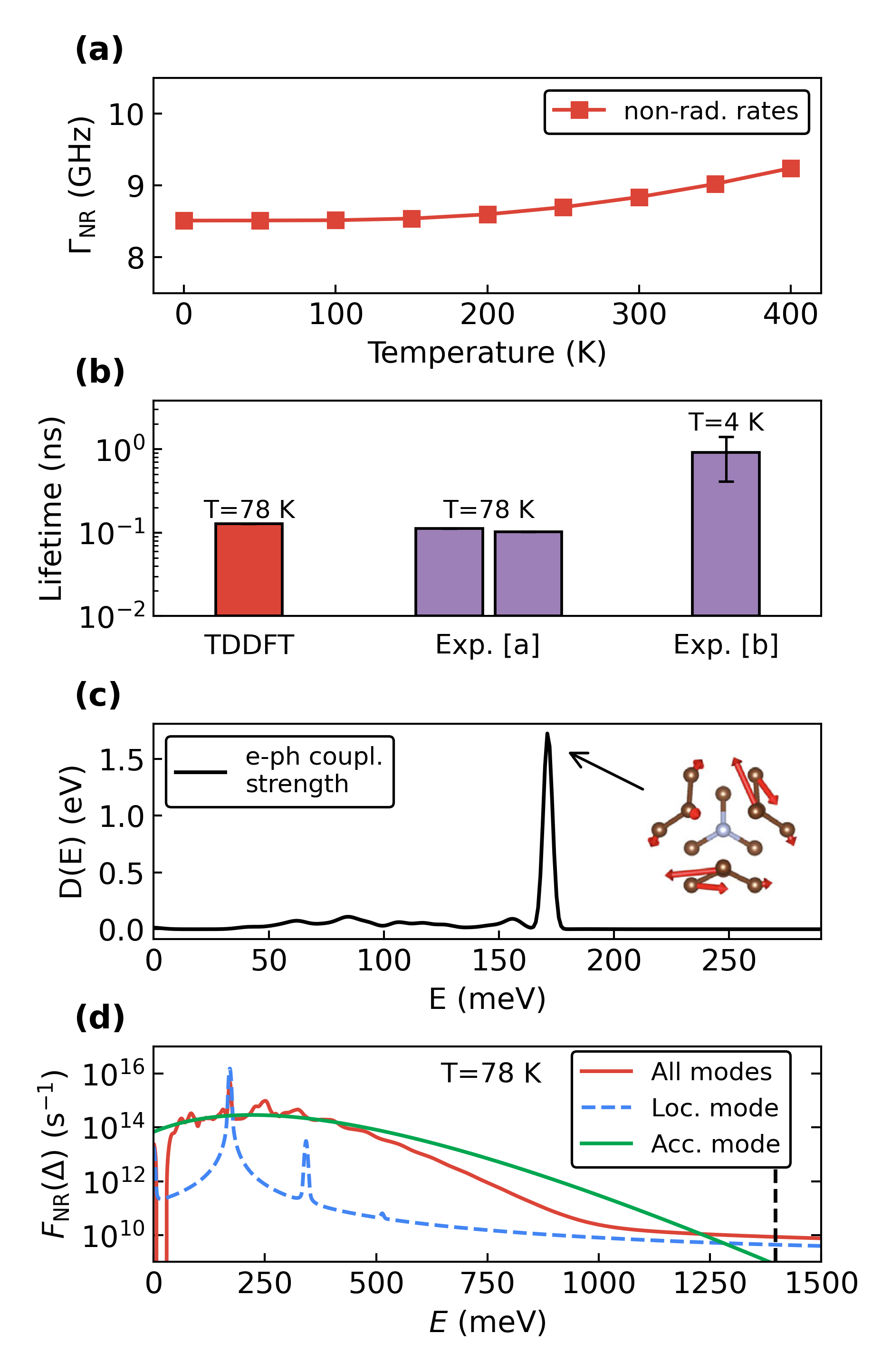}
    \end{minipage}
    \caption{\textbf{(a)}: Non-radiative transition rates between the singlet states in the NV$^{-}$ center in diamond (from Eq. (\ref{eq:def_NRTR})). \textbf{(b)}: Computed lifetimes of the ${}^1\mathrm{A}_1$ state compared to the experimental results of Ulbricht \textit{et al.}\cite{ulbricht2018excited} [a], and Acosta \textit{et al.}\cite{acosta2010optical}. [b]. \textbf{(c)}: Computed electron-phonon spectral density function [Eq. (\ref{eq:def_D(E)})], exhibiting a peak at the energy $E\simeq170~\mathrm{meV}$ of the vibrational mode localized around the defect (shown in the inset). \textbf{(d)}: Comparison between line-shape functions obtained within different approximations; the black dashed line indicates the TDDFT adiabatic energy difference $\sim\hspace{-0.11cm}1.397~\mathrm{eV}$.}
    \label{fig:singlets_NV}
\end{figure}

\textit{Results}---We first consider the IC processes between the singlet states of the NV$^-$ center in diamond. Using a recently developed spin-flip TDDFT framework\cite{jin2023excited}, we are able to capture the multi-configurational character of these states, absent in Kohn-Sham orbital descriptions. With a zero-phonon line (ZPL) in the infrared (IR) at 
% $\sim\hspace{-.11cm}1042$ nm 
$\sim\hspace{-0.11cm}1.19$ eV\cite{acosta2010optical,ulbricht2018excited}, the  $^1\mathrm{A}_1\to{}^1\mathrm{E}$ transition is predominantly non-radiative, as indicated by the measured integrated IR emission intensities, which are three orders of magnitude lower than those of the visible $^3\mathrm{E} \to {}^3\mathrm{A}_2$ emission\cite{acosta2010optical}. To shed light on the role of IC processes in the non-radiative decay channel, we computed the NRTRs presented in Figure \ref{fig:singlets_NV}\textbf{(a)}. We used a $215$-atom supercell, 
% $\gamma=0.18~\mathrm{meV}$ in Eq. (\ref{eq:def_G(E)}), following Ref. \cite{biktagirov2017strain}, 
and the adiabatic energy difference $\Delta_{IF}=1.397~\mathrm{eV}$ obtained with TDDFT and the DDH functional\cite{jin2023excited}. Calculations performed with the PBE functional yield qualitatively similar results (See SM).
% the experimental value of the ZPL $\sim\hspace{-.11cm}1.19$ eV as the adiabatic energy difference. 
In the range $T=0~\sim400~\mathrm{K}$ we find rates of the order of $\approx\hspace{-0.1cm}10^1~\mathrm{GHz}$. Since the predicted radiative rate is $\Gamma_\mathrm{R}\simeq0.53~\mathrm{MHz}$\cite{bockstedte2018ab}, our results indicate a predominant role of IC processes in the decay of the ${}^1\mathrm{A}_1$ state. 

\begin{figure}[!h]
    \begin{minipage}[c]{1.0\linewidth}
    \centering
    \includegraphics[width=1.0\textwidth]{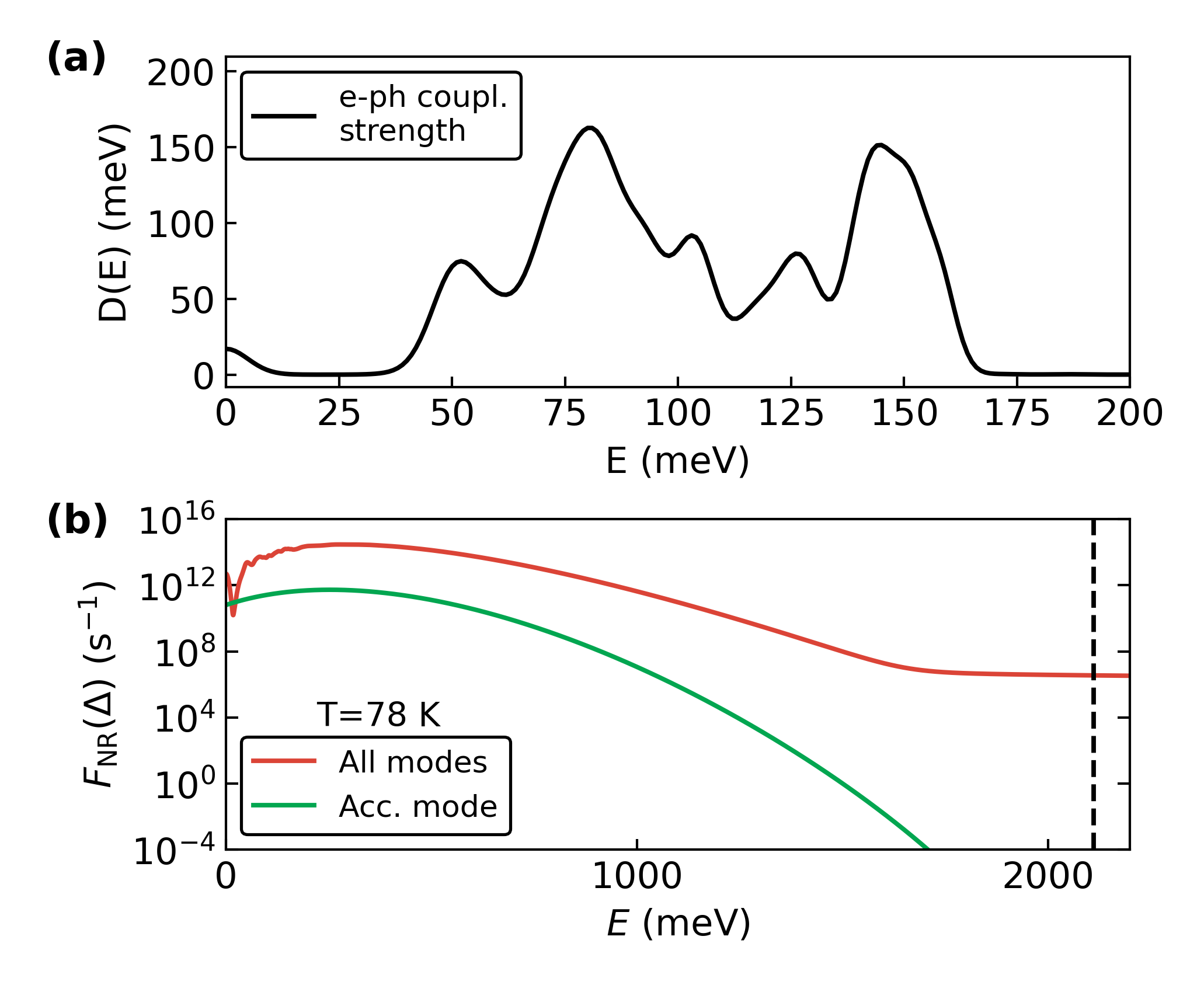}
    \end{minipage}
    \caption{\textbf{(a)}: Computed electron-phonon spectral density function $D(E)$ [Eq. (\ref{eq:def_D(E)})] for the transition between the triplet states of the NV${^-}$ center in diamond. \textbf{(b)}: Comparison between line-shape functions obtained within different approximations. The black dashed line indicates the TDDFT adiabatic energy difference $\Delta_{IF}=2.112~\mathrm{eV}$.}
    \label{fig:triplets_NV}
\end{figure}

\begin{figure}[!h]
    \begin{minipage}[c]{1.0\linewidth}
    \centering
    \includegraphics[width=1.0\textwidth]{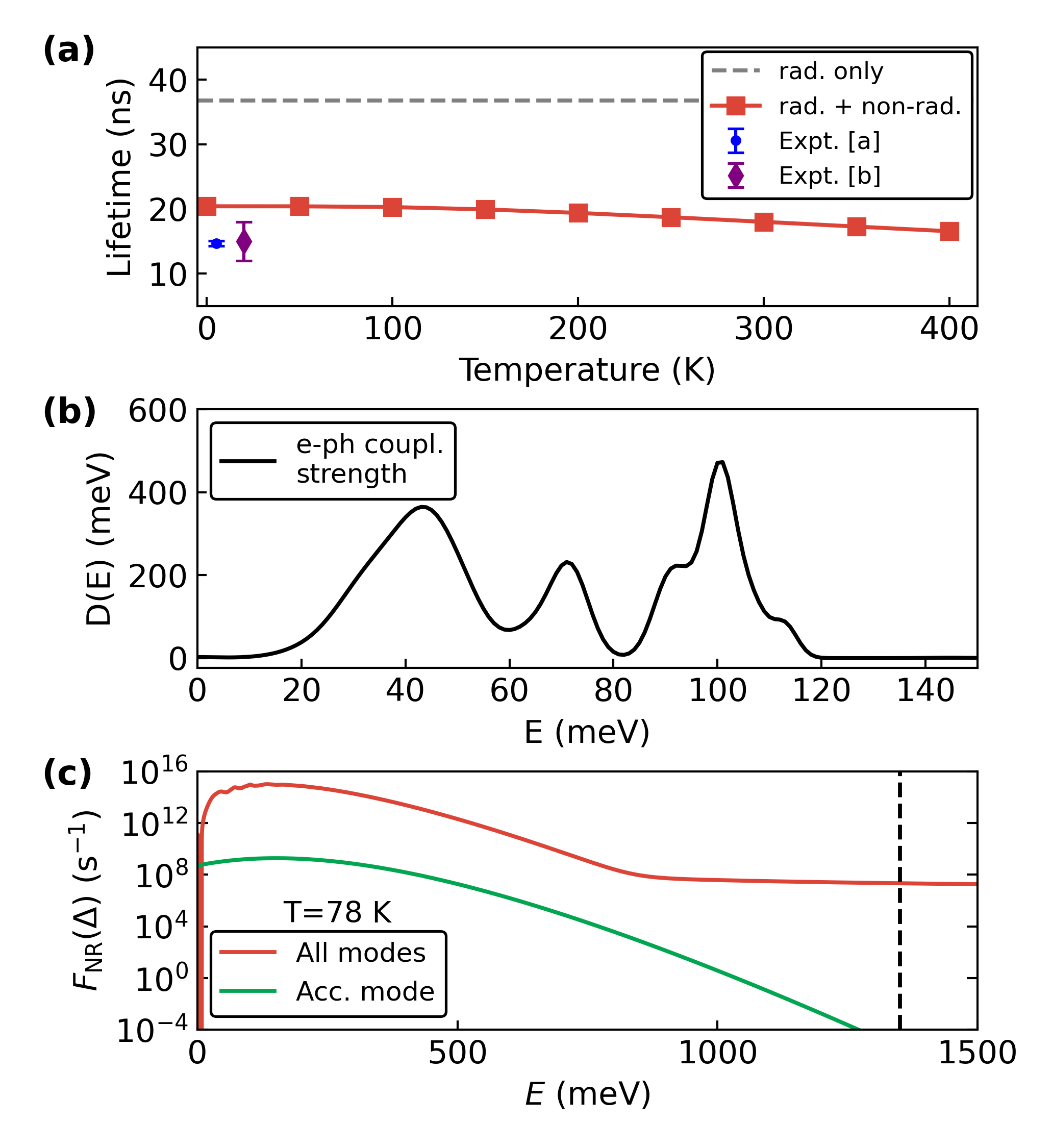}
    \end{minipage}
    \caption{\textbf{(a)}: Radiative lifetime of the ${}^3\mathrm{E}$ excited state of the \textit{kk}-VV$^0$ center in 4H-SiC\cite{jin2021photoluminescence} (gray dashed line), computed at the DFT level of theory with the PBE functional. The red squares show the total computed lifetime, including non-radiative decays, and the blue and purple dots indicate experimental measurements from Ref.s\cite{anderson2019electrical} [a] and \cite{falk2014electrically} [b]. \textbf{(b)}: Computed electron-phonon spectral density function $D(E)$ [Eq. (\ref{eq:def_D(E)})]. \textbf{(c)}: Comparison between line-shape functions obtained within different approximations. The black dashed line indicates the TDDFT adiabatic energy difference $\Delta_{IF}=1.351~\mathrm{eV}$.}
    \label{fig:triplets_SiC}
\end{figure}

We compared the lifetime $\tau^{{}^1\mathrm{A}_1}$ of the upper singlet state with experiments from Ref.s \cite{acosta2010optical} and \cite{ulbricht2018excited}, as shown in Figure \ref{fig:singlets_NV}\textbf{(b)}. Acosta \textit{et al.} carried out fluorescence lifetime measurements using the phase-shift technique, where a sinusoidal modulation of the intensity of an optical pump beam results in oscillations of the time-dependent IR fluorescence. From the relative phase shift between the pump and the response, they estimated  $\tau^{{}^1\mathrm{A}_1}=0.9(5)~\mathrm{ns}$ at $T=4~\mathrm{K}$, an order of magnitude smaller than the one evaluated here. We note, however, the large relative error of the  phase-shift estimate in the experiments of Acosta \textit{et al.}, which might be responsible for an inaccurate result. Ulbricht \textit{et al.} performed femtosecond transient absorption pump-probe spectroscopy, using a femtosecond optical pump to activate the optical cycle and a femtosecond probe resonant with the transition between the singlet states to study the population dynamics. From the time-dependent normalized differential transmission measured in two different samples, they inferred lifetimes of $\tau^{{}^1\mathrm{A}_1}=102.3(3)~\mathrm{ps}$ and $\tau^{{}^1\mathrm{A}_1}=92.1(5)~\mathrm{ps}$, at $T=78~\mathrm{K}$, in accord with our results ($\tau^{{}^1\mathrm{A}_1}=\Gamma_\mathrm{NR}{}^{-1}=117.4~\mathrm{ps}$ at $T=78~\mathrm{K}$). This good agreement further confirms that IC processes are the main cause of non-radiative decays for the $^1\mathrm{A}_1\to{}^1\mathrm{E}$ transition. 
\\ \indent 
It is interesting to define and analyze the e--ph spectral density function 
\begin{equation}
    \label{eq:def_D(E)}
    D(E) = \frac{1}{2\pi\hbar} \sum_k^{N_\mathrm{modes}} |M_{IF}^{k}|^2 B_k(E)
\end{equation}
% is the spectral density for the electron-phonon coupling, 
to identify the role played by various phonon modes in the IC process. The vibrational mode localized around the defect at $E\simeq170~\mathrm{meV}$, shown in Figure \ref{fig:singlets_NV}\textbf{(c)}, has a strong non-adiabatic character, and it might be expected to be the main driver of IC processes. However, when computing the NRTRs by considering only this mode in the sum of Eq. (\ref{eq:def_D(E)}), we obtain $\Gamma_{\mathrm{NR,loc}}=4.37~\mathrm{GHz}$ at $T=78~\mathrm{K}$, which is roughly half the rate computed with the inclusion of all the modes ($\Gamma_{\mathrm{NR}}=8.52~\mathrm{GHz}$). This comparison highlights the importance of including all phonon modes to obtain accurate results, and not only those localized around the defect. 
\\ \indent
Our framework represents a substantial improvement over commonly used methods in which, to reduce the computational cost of Eq. (\ref{eq:def_NRTR}) two approximations to $X_{IF}^k$ are usually adopted: the single-mode approximation\cite{kazemi2026giant}, where only one of the $N_\mathrm{modes}$ modes is considered, or the accepting-mode approximation\cite{turiansky2021nonrad}, where an effective mode $Q$ is constructed from the mass-weighted displacement between the initial $\boldsymbol{R}_I$ and the final $\boldsymbol{R}_F$ atomic configurations 
% i.e., $Q^2=\sum_AM_A(R_I-R_F)^2$, 
and the vibrational energies are extracted from the one-dimensional configuration-coordinate diagrams of the electronic energies with respect to $Q$. In Table \ref{tab:NRTR_singlets_NV} we compare the NRTRs computed with different approaches, and in Figure \ref{fig:singlets_NV}\textbf{(d)} we show the line-shape functions obtained with the different approximations. Compared to the all-modes line-shape function, the single-mode approximation describes better the tail of the line-shape, whereas the accepting-mode approximation performs better at lower energies. We also compare the NRTR at $T=300~\mathrm{K}$ computed in the accepting-mode approximation $\Gamma_{\mathrm{NR,acc.}}=2.49~\mathrm{GHz}$ against the value $\Gamma_{\mathrm{NR,acc.}}=0.048~\mathrm{GHz}$ obtained using Kohn-Sham electronic orbitals in Ref.~\cite{li2024excited}. The orders of magnitude difference between the two approaches highlights the importance of an accurate description of the many-body character of the electronic states. (See Appendix).
\begin{table}[!htb]
\centering
\caption{Comparison of non-radiative transition rates between the singlet states of the NV${^-}$ center in diamond obtained from first-principles with different approximations, compared with experimental results.}
\label{tab:NRTR_singlets_NV}
\begin{tabular}{cccc}
\toprule
\toprule
% \hline \hline
& \multicolumn{3}{c}{$\Gamma_{\mathrm{NR}}\,[{}^1\mathrm{A}_1 \to {}^1\mathrm{E}]$  (GHz)} \\
% \cmidrule(lr)
\addlinespace[2pt]
 & Acc. mode & Loc. modes & All modes \\
\midrule
& \multicolumn{3}{c}{$T = 4\,\mathrm{K}$} \\
\addlinespace[2pt]
% \makecell[c]{TDDFT (PBE)} & $54.44$ & $4.610$ & $8.275$ \\
\makecell[c]{TDDFT} & $3.27$ & $4.37$ & $8.51$ \\
% \makecell[c]{DFT\tablenote{Ref.\cite{li2024excited}} (DDH)} & $0.0114$ & -- & -- \\
\makecell[c]{Exp.\tablenote{Ref.\cite{acosta2010optical}}} & -- & -- & $1.1(6)$ \\
\midrule
& \multicolumn{3}{c}{$T = 78\,\mathrm{K}$} \\
\addlinespace[2pt]
% \makecell[c]{TDDFT (PBE)} & $54.45$ & $4.610$ & $8.279$ \\
\makecell[c]{TDDFT} & $3.27$ & $4.37$ & $8.52$ \\
% \makecell[c]{DFT\tablenote{Ref.\cite{li2024excited}} (DDH)} & $0.0158$ & -- & -- \\
\makecell[c]{Exp.\tablenote{Ref.\cite{ulbricht2018excited}}} & -- & -- & \makecell[c]{10.858(59) \\ 9.775(29)} \\
% \midrule
% & \multicolumn{3}{c}{$T = 300\,\mathrm{K}$} \\
% \addlinespace[2pt]
% \makecell[c]{TDDFT (PBE)} & $135.14$ & $4.620$ & $8.60$ \\
% \makecell[c]{TDDFT (DDH)} & $49.15$ & $5.728$ & $12.27$ \\
% \makecell[c]{DFT\tablenote{Ref.\cite{li2024excited}} (DDH)} & $0.0482$ & -- & -- \\
\bottomrule
\bottomrule
\end{tabular}
\end{table}
\\ \indent 
For completeness, we also considered the ${}^3\mathrm{E}\to{}^3\mathrm{A}_2$ transition in NV$^-$, which is known to be predominantly radiative, with an experimental ZPL $\sim\hspace{-0.11cm}1.95~\mathrm{eV}$. Recent first-principles estimates of the lifetime $\tau^{{}^3\mathrm{E}}$ of the triplet excited state including radiative and  non-radiative ISC contributions, agree well with experimental measurements\cite{jin2023excited,jin2025first}, suggesting a negligible role of IC processes. Consistent with this interpretation, here we predict $\Gamma_{\mathrm{NR}}\lesssim 5~\mathrm{MHz}$ in the range $T=0\sim400~\mathrm{K}$, a negligible contribution compared to the radiative rate $\Gamma_{\mathrm{R}}\simeq83.3~\mathrm{MHz}$\cite{jin2025first}. This finding is not surprising since the ground and excited triplet states are well separated in energy. Our results were obtained 
% by setting $\gamma=2.07\times 10^{-4}~\mathrm{meV}$ in Eq. (\ref{eq:def_G(E)}), estimated from the low-temperature photoluminescence linewidths measured in Ref. \cite{fu2009observation}, and 
considering  
% ZPL $\sim\hspace{-0.11cm}1.95~\mathrm{eV}$
the adiabatic energy difference $\Delta_{IF}=2.112~\mathrm{eV}$, as obtained with TDDFT, and the DDH functional. In Figure \ref{fig:triplets_NV}\textbf{(a)} we show the e--ph spectral function $D(E)$. We do not observe any preferential contribution from any specific mode, suggesting that single-mode approximations may not be valid. Indeed, the line-shape function obtained within the accepting-mode approximation, Figure \ref{fig:triplets_NV}\textbf{(b)}, underestimates by several order of magnitude the rates obtained when including all phonons. 
\\ \indent 
We now turn to IC processes occurring in the optical cycle of the \textit{kk}-VV$^0$ center in the 4H-SiC polytype\cite{falk2013polytype}. Although non-radiative decays are expected to play an important role\cite{bockstedte2018ab}, a thorough understanding of the contribution of IC processes has been hampered by the lack of an adequate computational framework. We first consider the ${}^3\mathrm{E}\to{}^3\mathrm{A}_2$ transition. The energy separation between the triplet ground and excited states (with a ZPL $\sim\hspace{-0.1cm}1.096~\mathrm{eV}$) has often been considered to be sufficiently large to render non-radiative contributions from IC processes negligible. Nevertheless, state-of-the-art first-principles estimates of the lifetime of the ${}^3\mathrm{E}$ state obtained by considering only the radiative contribution overestimate the experimental measurements of $\tau^{{}^3\mathrm{E}}\simeq15~\mathrm{ns}$\cite{falk2014electrically,anderson2019electrical}, when using either DFT ($\tau^{{}^3\mathrm{E}}_{\mathrm{R}}\simeq36.8~\mathrm{ns}$)\cite{jin2021photoluminescence} or constrained DFT ($\tau^{{}^3\mathrm{E}}_{\mathrm{R}}\simeq28~\mathrm{ns}$)\cite{bian2025theory}. Here, we include the contributions of IC processes using the relation 
\begin{equation}
    \tau^{{}^3\mathrm{E}}= \left[ \frac{1}{\tau^{{}^3\mathrm{E}}_{\mathrm{R}}}\,+\,\Gamma_{\mathrm{NR}}\;[{}^3\mathrm{E}\to{}^3\mathrm{A}_2] \right]^{-1}.
\end{equation}
We considered a $398$-atom supercell, 
% $\gamma=4.13\times10^{-4}~\mathrm{meV}$, as estimated from low temperature photoluminescence linewidths\cite{christle2017isolated}, 
we used the DDH functional and computed
% the ZPL $\sim\hspace{-0.11cm}1.096~\mathrm{eV}$ as 
the adiabatic energy difference $\Delta_{IF}=1.351~\mathrm{eV}$\cite{jin2023excited} with TDDFT. (A comparison with PBE results is given in the SM). We present our results in Figure \ref{fig:triplets_SiC}\textbf{(a)}, where we show that the inclusion of IC processes improves the agreement with experiments. In Figure \ref{fig:triplets_SiC}\textbf{(b)} we display the e--ph spectral function $D(E)$, and in Figure \ref{fig:triplets_SiC}\textbf{(c)} we analyze the line-shape functions, comparing with the results obtained with the accepting-mode approximation. Similar to the decay between triplet states in the NV$^-$, we observe no predominant phonon mode contribution; indeed, the effective mode approximation severely underestimates the NRTRs, highlighting once again the importance of an approach that includes all-phonon-modes. 
\\ \indent
Finally, we analyze the $^1\mathrm{A}_1\to{}^1\mathrm{E}$ transition, which has not yet been investigated experimentally and for which neither a ZPL estimate nor singlet-state lifetime measurements are available. Hence our results serve as predictions. We computed the NRTRs using the adiabatic energy difference $\Delta_{IF}=1.132~\mathrm{eV}$ obtained from TDDFT\cite{jin2023excited}. (See a comparison with PBE results in the SM). We 
% use the same $\gamma=0.18~\mathrm{meV}$ as adopted for the singlet states of the NV$^-$ and
obtain rates of the order of $\approx10^1~$GHz, similar to those computed for the NV${^-}$; comparing our results with the first-principles value of the radiative rate $\Gamma_\mathrm{R}=0.32~\mathrm{MHz}$\cite{bockstedte2018ab}, we conclude that the $^1\mathrm{A}_1\to{}^1\mathrm{E}$ transition is dominated by IC processes.

\textit{Discussion}---We have introduced a general, efficient first-principles framework for computing non-radiative spin-conserving transition rates in spin defects, addressing a longstanding gap in the theoretical description of the optical cycle of these systems. Two advances are central to our approach. First, we treat the electronic states of defects as many-body wavefunctions computed within LR-TDDFT with hybrid functionals; hence we capture multi-configurational effects that single-particle Kohn-Sham orbitals cannot describe; indeed, using Kohn-Sham single-particle wavefunctions underestimates rates by orders of magnitude. Second, we compute non-adiabatic coupling vectors analytically using an extended Lagrangian formulation; we thus eliminate the bottleneck of finite-difference calculations of wave-function derivatives, and enable the inclusion of all phonon modes for supercells with hundreds of atoms. Applying our framework to the NV$^-$ center in diamond, we find IC rates for the singlet ${}^1\mathrm{A}_1\to{}^1\mathrm{E}$ transition in quantitative agreement with femtosecond transient absorption measurements, and confirm the negligible role of IC in the ${}^3\mathrm{E}\to{}^3\mathrm{A}_2$ decay. For the VV$^0$ center in 4H-SiC, we identify a significant non-radiative contribution to the ${}^3\mathrm{E}$ lifetime that has been previously overlooked, resolving a discrepancy between computed and measured lifetimes present in the literature. For the singlet ${}^1\mathrm{A}_1\to{}^1\mathrm{E}$ transition in VV$^0$, where no experimental data exist, our calculations serve as quantitative predictions. For defects in both diamond and SiC, we demonstrate that effective single-mode and accepting-mode approximations underestimate non-radiative rates by orders of magnitude.
\\ \indent
Together with first-principles methods for inter-system crossing, our framework enables parameter-free predictions of complete optical cycles and ODMR spectra of spin defects. Our approach is broadly applicable to non-adiabatic processes in heterogeneous solids and molecules, and the analytical NAC implementation paves the way for ab initio non-adiabatic molecular dynamics in extended systems and the study of polaron and exciton dynamics in solids. Work is in progress to further reduce computational cost using machine-learned dynamical matrices and NAC surrogate models.

\textit{Acknowledgments}---We gratefully acknowledge fruitful discussions with Chris Anderson, Enrico Drigo, and Michael Nevins. The computational work was supported by the Midwest Integrated Center for Computational Materials (MICCoM) as part of the Computational Materials Sciences Program funded by the U.S. Department of Energy and the study of spin defects was supported by AFOSR grant \# FA95502210370. This research used resources of the National Energy Research Scientific Computing Center (NERSC), a DOE Office of Science User Facility supported by the Office of Science of the U.S. Department of Energy under Contract No. ERCAP0036175 and resources of the University of Chicago Research Computing Center (RCC). The Flatiron Institute is a division of the Simons Foundation.

\textit{Data availability}---The code, inputs and data that support the findings of this article are openly available through the Qresp platform\footnote{\texttt{https://paperstack.uchicago.edu/}}

\bibliographystyle{unsrt} % BibTeX style
\bibliography{bibliography} 

\appendix

\begin{center}
\textbf{End Matter}
\end{center}

% \section{Appendix I: Generating function approach}
% \label{sec:app_gen_func}
\textit{Appendix I: Generating function approach}---Here, we provide the expressions for the generating functions $G(E)$ and $B_k(E)$ entering Eq. (\ref{eq:def_NRTR}). They are derived from the phonon-relaxation term, Eq. (\ref{eq:def_Xif}), in the displaced harmonic oscillator approximation, and they describe the multi-phonon processes at temperature $T$, that account for the conservation of the total energy and for the overlap between Franck-Condon-shifted phononic wavefunctions\cite{nichols2025multiphonon}. 
The generating function $G(E)$ reads
\begin{equation} 
    \label{eq:def_G(E)}
    G(E) = \frac{1}{2\pi \hbar} \int_{-\infty}^{\infty} G(t) e^{\frac{iEt - \gamma |t|}{\hbar}} dt
\end{equation}
where $\gamma$ is a Lorentzian smearing factor accounting for the lifetime broadening of the electronic levels\cite{nichols2025multiphonon}, and it holds
\begin{equation}
\begin{aligned}
    G(t) = \prod_{k=1}^{N_\mathrm{modes}} G_k (t) = \exp & \biggl\{ \sum_k^{N_\mathrm{modes}} S_k \left[ \left( \overline{n}_k + 1 \right) e^{i\omega_k t} +\right. \\ &\quad \left. + \overline{n}_k e^{-i\omega_k t} - \left( 2 \overline{n}_k + 1 \right)   \right] \biggr\}
\end{aligned}
\end{equation}
where $\hbar\omega_k$ is the energy of the $k$-th mode, $\overline{n}_k=(e^{\hbar\omega_k/k_\mathrm{B}T}-1)^{-1}$ is the thermal occupation at temperature $T$, $k_\mathrm{B}$ is the Boltzmann constant, and $S_k$ is the partial Huang-Rhys factor\cite{huang1950theory}, namely
\begin{equation}
    S_k = \frac{\omega_k\Delta Q_k^2}{2\hbar}
\end{equation}
with 
\begin{equation}
    % \Delta Q_k = \sum_{A=1}^{N_\mathrm{atoms}}\sqrt{M_A}(\bs{R}_{I,A}-\bs{R}_{F,A})\cdot\bs{\mathrm{e}}^k_A \;.
    \Delta Q_k = \frac{1}{\omega_k^2}\sum_{A=1}^{N_\mathrm{atoms}} \frac{\bs{\mathrm{F}}_{A}}{\sqrt{M_A}}\cdot\bs{\mathrm{e}}^k_A \,,
\end{equation}
where $\bs{\mathrm{F}}$ is the final-state (excited-state) force evaluated at the initial state relaxed geometry.
The function $B_k(E)$, instead, reads: 
\begin{equation}
    B_k(E) = \int_{-\infty}^{\infty} B_k(t) e^{\frac{iEt}{\hbar}} \dd t
\end{equation}
where
\begin{equation}
\begin{aligned} 
    B_k(t) &= \frac{\hbar}{2\omega_k} \Bigl\{ e^{i\omega_k t} + 2\overline{n}_k \cos(\omega_k t)+ \\
    &\qquad + S_k \left[ - e^{i\omega_k t} - 2i\overline{n}_k \sin(\omega_k t) + 1 \right]^2 \Bigr\}.
\end{aligned} 
\end{equation} 

% \section{Appendix II: Effective-modes approximations}
\textit{Appendix II: Effective-modes approximations---}We provide below details on our calculations of NRTRs in the accepting-mode approximation. Following a standard procedure, we defined an effective mode $Q$ as the mass-weighted atomic displacement from the relaxed geometry of the initial state $\bs{R}_I$ to that of the final state $\bs{R}_F$, namely
% \begin{equation}
%     Q = \left[\sum_A^{N_\mathrm{atoms}}\sum_{i=x,y,z} M_A ( R_{i,A}-R_{F;iA} )^2 \right]^{1/2} \,,
% \end{equation}
\begin{equation}
    Q^2 = \sum_A^{N_\mathrm{atoms}}\sum_{i=x,y,z} M_A ( R_{i,A} - R_{I;iA} )^2 \,,
\end{equation}
where the displacement $R_{i,A}$ of atom $A$ along the Cartesian direction $i$ is proportional to $R_{I;iA}-R_{F;iA}$. We defined the non-radiative rate $\Gamma_{\mathrm{NR},\mathrm{acc.}}(T)$ as
\begin{equation}
    \label{eq:def_NRTR_acc}
    \Gamma_{\mathrm{NR},\mathrm{acc.}}(T) = \frac{2\pi}{\hbar} |M_{IF}|^2 X_{IF}(T) \,.
\end{equation}
The e--ph coupling term $M_{IF}$ reads 
\begin{align}
    \label{eq:def_Mif_1D}
    M_{IF}
    &= \langle \Psi_I| \frac{\partial H_\mathrm{el} }{\partial Q} | \Psi_F \rangle \,, \\
    \label{eq:def_Mif_acc_mode_dwfc}
    &= (E_F - E_I) \langle \Psi_I| \frac{\partial \Psi_F }{\partial Q} \rangle \,, \\
    \label{eq:def_Mif_acc_mode_NACs}
    &= (E_F - E_I) \sum_A^{N_\mathrm{atoms}} \bs{d}_{IF,A} \cdot 
   \frac{\hat{\bs{r}}_{IF,A}}{\sqrt{M_A}}
\end{align}
where we projected the NAC vectors along the direction $\hat{\bs{r}}_{IF}$ of the atomic displacement from $\bs{R}_I$ to $\bs{R}_F$, namely
\begin{equation}
    \hat{\bs{r}}_{IF} = \frac{\bs{R}_{I}-\bs{R}_{F}}{|\bs{R}_{I}-\bs{R}_{F}|} \,.
\end{equation} 
The phonon-relaxation term $X_{IF}(T)$ reads
\begin{equation}
    \label{eq:def_Xif_1D}
    X_{IF}(T) = \sum_{m,n} \rho_n(T) | \langle \Phi_{I,n} | Q | \Phi_{F,m} \rangle |^2 \times \delta(E_{I,n}^{\mathrm{tot}}  - E_{F,m}^{\mathrm{tot}}) \,,
\end{equation}
where $n$ and $m$ are the occupation numbers of the initial and final effective-mode vibrational states, respectively, and 
\begin{equation}
\begin{aligned}
    E_{I,n}^{\mathrm{tot}}  - E_{F,m}^{\mathrm{tot}} = \Delta_{IF} + \left(n+\frac{1}{2}\right)\hbar\omega_I - \left(m+\frac{1}{2}\right)\hbar\omega_F \,,
\end{aligned}
\end{equation}
% \begin{align}
%     E_{L,l}^{\mathrm{tot}} &= E_L(\bs{R}_L) + \left(l+\frac{1}{2}\right)\hbar\omega_{L} \,,
% \end{align}
% with $L,l=I,n$ or $L,l=F,m$. 
The vibrational energies $\hbar\omega_I$ and $\hbar\omega_F$ are extracted from the configuration-coordinate diagram of the electronic energies $E_I$ and $E_F$ as a function of $Q$, respectively. 
\\ \indent
We used the accepting-mode approximation to compute the NRTRs of the IC processes occurring in the NV$^-$ in diamond and in the VV$^0$ in 4H-SiC, with the parameters given in Table \ref{tab:acc_mode_parameters}, which were obtained with TDDFT and the DDH functional. For the ${}^3\mathrm{E}\to{}^3\mathrm{A}_2$ transitions, we extracted the values of $\hbar\omega_I$ and $\hbar\omega_F$ from the configuration-coordinate energy diagrams obtained with the DDH functional.
% and displayed in Figure \ref{fig:1Dcc_diagrams}. 
For the ${}^1\mathrm{A}_1\to{}^1\mathrm{E}$ transition of the NV$^-$ in diamond, we used the same values adopted in Ref.~\cite{li2024excited} to ensure a consistent comparison of the rates; the important difference is that our rates are obtained using the NACs to compute the e--ph coupling term instead of finite differences of Kohn-Sham orbitals. 

\begin{table}[!htb]
\centering
\caption{Parameters used to compute the NRTRs in the accepting-mode approximation.}
\vspace{0.2cm}
\label{tab:acc_mode_parameters}
\begin{tabular}{lccc}
\hline
\hline
\addlinespace[2pt]
  & $\Delta Q$ (\AA~amu$^{1/2}$) & $\hbar\omega_I$ (meV) & $\hbar\omega_F$ (meV) \\
\midrule
\addlinespace[2pt]
${}^3\mathrm{E}\to{}^3\mathrm{A}_2$ in NV$^-$ & 0.63 & $72.96$ & $66.54$ \\
${}^3\mathrm{E}\to{}^3\mathrm{A}_2$ in \textit{kk}-VV$^0$ & 0.78 & $36.41$ & $38.70$ \\
${}^1\mathrm{A}_1\to{}^1\mathrm{E}$ in NV$^-$\tablenote{Values taken from Ref.~\cite{li2024excited}.} & 0.42 & $74.07$ & $87.34$ \\
\hline
\hline
\end{tabular}
\end{table}
% \begin{figure}[!h]
%     \begin{minipage}[c]{1.0\linewidth}
%     \centering
%     \includegraphics[width=1.0\textwidth]{Figures/NV_En_vs_q_1dcc.png} \\
%     \includegraphics[width=1.0\textwidth]{Figures/SiC_En_vs_q_1dcc.png}
%     \end{minipage}
%     \caption{\textbf{(a)}: Configuration-coordinate diagrams of the ${}^3\mathrm{A}_2$ and ${}^3\mathrm{E}$ states of the NV$^-$ center in diamond. \textbf{(b)}: Configuration-coordinate diagrams of the ${}^3\mathrm{A}_2$ and ${}^3\mathrm{E}$ states of the \textit{kk}-VV$^0$ center in 4H-SiC. In both panels, the vertical dashed lines indicate the position of the minima.}
%     \label{fig:1Dcc_diagrams}
% \end{figure}
We emphasize that the use of NACs in Eq. \ref{eq:def_Mif_acc_mode_NACs} already constitutes a fundamental advancement over common practices, even in the effective-mode approximation. Indeed, it eliminates the need to compute wave-function derivatives with finite differences, and allows for the inclusion of multi-configurational effects in the description of the electronic states, which is missing in single-particle approaches. In Table \ref{tab:Mif_comparison} we compare the values of $M_{IF}$ computed with NACs against those obtained using Kohn-Sham electronic orbitals. The latter are obtained by approximating the expression in Eq. (\ref{eq:def_Mif_acc_mode_dwfc}) as follows
\begin{equation}
    (E_F - E_I) \langle \Psi_I| \frac{\partial \Psi_F }{\partial Q} \rangle \approx (\varepsilon_F-\varepsilon_I) \langle \varphi_I|\frac{\partial\varphi_F}{\partial Q} \rangle \,,
\end{equation}
where $\varphi_I$ ($\varphi_F$) and $\varepsilon_I$ ($\varepsilon_F$) indicate the Kohn-Sham electronic orbitals and energies of the initial (final) state, respectively. The overlap derivatives are computed by linear interpolation of the term $\langle \varphi_I (0)| \varphi_F(Q) \rangle$ with respect to $Q$. To approximate the electronic excited states with Kohn-Sham orbitals, we adopted the standard C$_{3v}$ point group minimal model for the defect's orbitals\cite{PhysRevB.98.085207}.
\begin{table}[!htb]
\centering
\caption{Values of the electron-phonon coupling term $M_{IF}$ computed in the accepting-mode approximation using NACs compared with values obtained using Kohn-Sham (KS) electronic orbitals.}
\vspace{0.2cm}
\label{tab:Mif_comparison}
\begin{tabular}{lcc}
\hline
\hline
\addlinespace[2pt]
  & \multicolumn{2}{c}{$M_{IF}$ (meV/\AA~amu$^{1/2}$)} \\
  & w/ NACs & w/ KS \\
\midrule
\addlinespace[2pt]
${}^3\mathrm{E}\to{}^3\mathrm{A}_2$ in NV$^-$ & $7.274$ & $0.387
$ \\
${}^3\mathrm{E}\to{}^3\mathrm{A}_2$ in \textit{kk}-VV$^0$ & $0.297$ & $0.142$ \\
${}^1\mathrm{A}_1\to{}^1\mathrm{E}$ in NV$^-$ & $239.2$ & $10.5$\tablenote{Ref.~\cite{li2024excited}.} \\
\hline
\hline
\end{tabular}
\end{table}

\end{document}

\setcounter{secnumdepth}{3}

\title{Supplemental Material for:\\
``First-principles calculations of internal conversion processes in spin defects''}

\author{Stefano Paolo Villani}
\email[]{svillani@uchicago.edu}
\affiliation{Pritzker School of Molecular Engineering, University of Chicago, Chicago, Illinois 60637, USA}

\author{Yu Jin}
\affiliation{Pritzker School of Molecular Engineering, University of Chicago, Chicago, Illinois 60637, USA}
\affiliation{Initiative for Computational Catalysis, Flatiron Institute, New York, NY 10010, USA}

\author{Giulia Galli}
\email[]{gagalli@uchicago.edu}
\affiliation{Pritzker School of Molecular Engineering, University of Chicago, Chicago, Illinois 60637, USA}
\affiliation{Department of Chemistry, University of Chicago, Chicago, Illinois 60637, USA}
\affiliation{Materials Science Division and Center for Molecular Engineering, Argonne National Laboratory, Lemont, Illinois 60439, USA}

% \date{\today}

\maketitle

\section{Comparison of non-radiative rates obtained with different DFT functionals}
\begin{figure}[!htb]
    \begin{minipage}[c]{1.0\linewidth}
    \centering
    \includegraphics[width=1.0\textwidth]{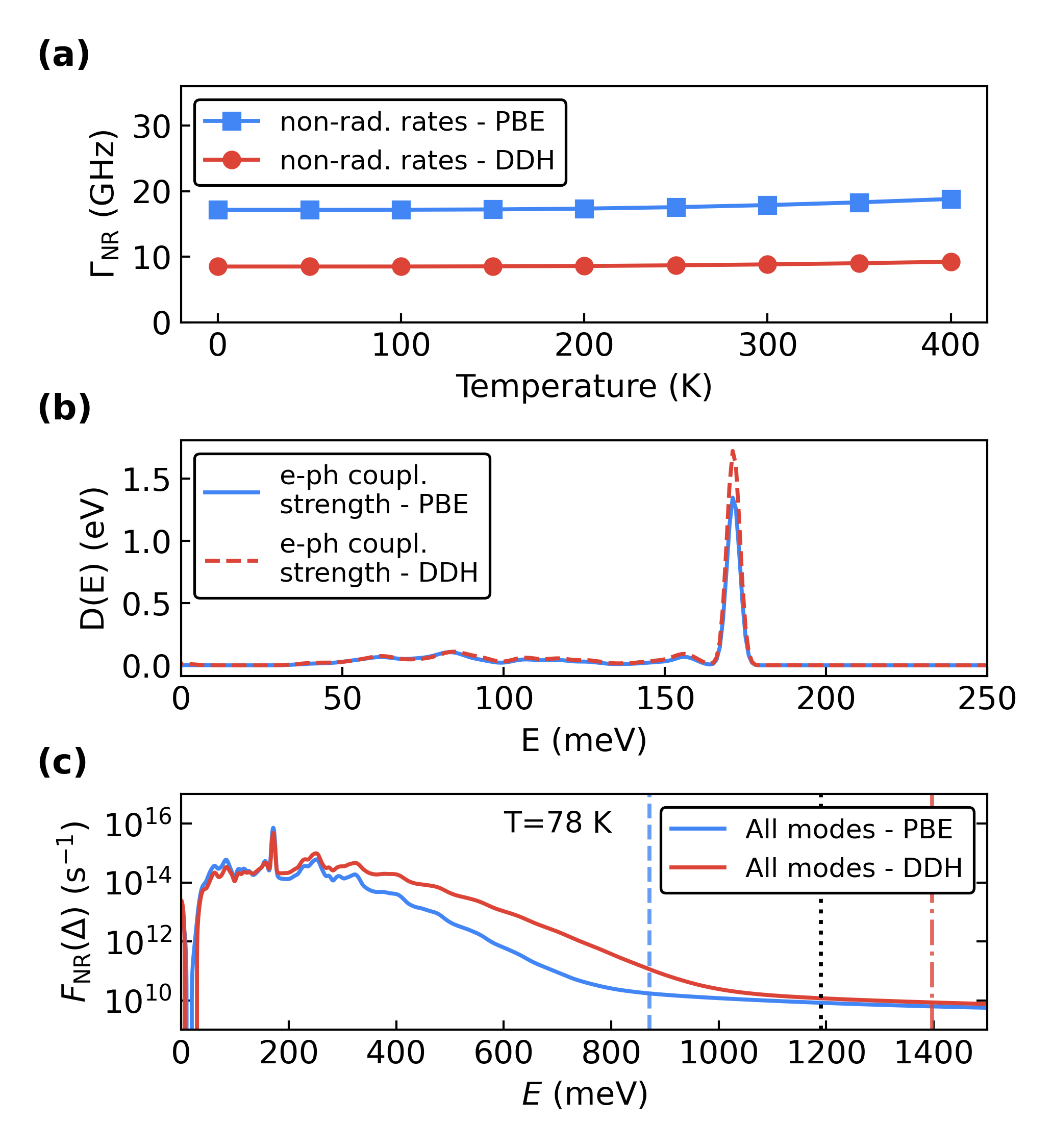}
    \end{minipage}
    \caption{\textbf{(a)}: Non-radiative rates of the ${}^1\mathrm{A}_1\to{}^1\mathrm{E}$ transition of the NV$^-$ in diamond for temperatures in the range $T=0\sim400~\mathrm{K}$ obtained using the PBE and DDH functionals. \textbf{(b)}: Comparison between the e-ph coupling spectral density functions. \textbf{(c)}: Comparison between the line-shape functions. The blue dashed line and the red dotted-dashed line indicate the adiabatic energy differences computed with TDDFT and the PBE and DDH functionals, respectively. The black dotted line indicates the experimental value of the ZPL.}
    \label{fig:NV_singlets_PBE_vs_DDH}
\end{figure}

\begin{figure}[!htb]
    \begin{minipage}[c]{1.0\linewidth}
    \centering
    \includegraphics[width=1.0\textwidth]{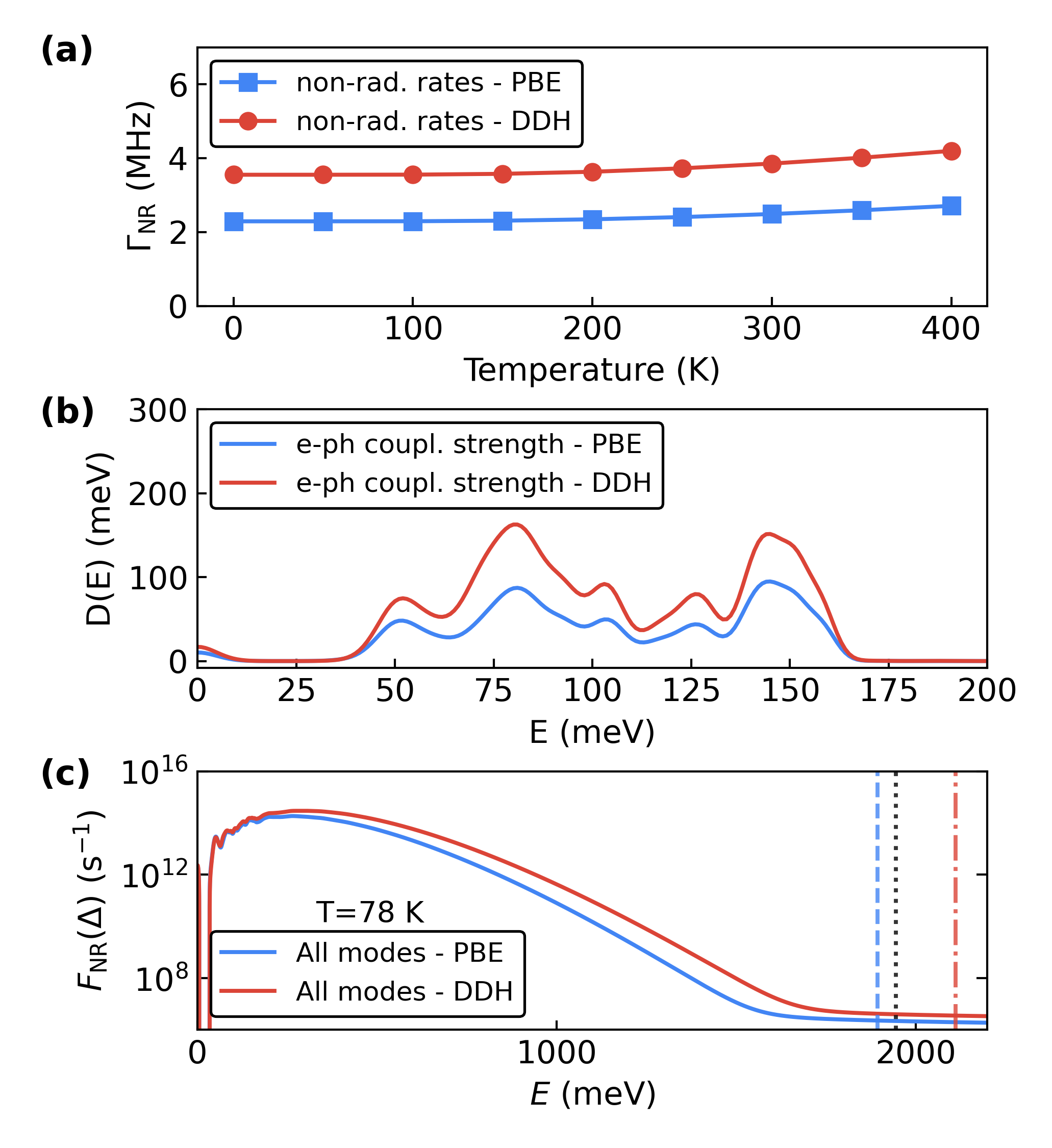}
    \end{minipage}
    \caption{\textbf{(a)}: Non-radiative rates of the ${}^3\mathrm{E}\to{}^3\mathrm{A}_2$ transition of the NV$^-$ in diamond for temperatures in the range $T=0\sim400~\mathrm{K}$ obtained using the PBE and DDH functionals. \textbf{(b)}: Comparison between the e-ph coupling spectral density functions. \textbf{(c)}: Comparison between the line-shape functions. The blue dashed line and the red dotted-dashed line indicate the adiabatic energy differences computed with TDDFT and the PBE and DDH functionals, respectively. The black dotted line indicates the experimental value of the ZPL.}
    \label{fig:NV_triplets_PBE_vs_DDH}
\end{figure}

Here, we analyze the effects of the DFT functional on the NRTRs by comparing results obtained with the PBE and DDH functionals. We start with the ${}^1\mathrm{A}_1\to{}^1\mathrm{E}$ transition of the NV$^-$ in diamond. In Figure \ref{fig:NV_singlets_PBE_vs_DDH}\textbf{(a)} we compare the NRTRs obtained with the two functionals and observe that the computed values differ by a factor of $\sim2$. By analyzing the e-ph spectral densities and the line-shape functions, displayed, respectively, in Figures \ref{fig:NV_singlets_PBE_vs_DDH}\textbf{(b)} and \ref{fig:NV_singlets_PBE_vs_DDH}\textbf{(c)}, we conclude that this discrepancy mainly arises from the large difference between the adiabatic energy differences computed with the two functionals ($\Delta_{IF}=0.871$ eV with PBE and $\Delta_{IF}=1.397$ eV with DDH). If we use the experimental ZPL value $1.195$ eV instead of the PBE or DDH adiabatic energy differences, the ratio between the values of the NRTRs is reduced to $\sim1.5$.
\clearpage
\newpage
Next, we analyze the ${}^3\mathrm{E}\to{}^3\mathrm{A}_2$ transition of the NV$^-$ in diamond. Similar to the ${}^1\mathrm{A}_1\to{}^1\mathrm{E}$ transition, the values of the NRTRs computed with the two functionals differ by a factor $\sim1.5$, as can be observed in Figure \ref{fig:NV_triplets_PBE_vs_DDH}\textbf{(a)}. However, the discrepancies now arise from sizable differences in both the e-ph coupling spectral functions and the adiabatic energy differences ($\Delta_{IF}=1.894$ eV with PBE and $\Delta_{IF}=2.112$ eV with DDH), as can be observed in Figures \ref{fig:NV_triplets_PBE_vs_DDH}\textbf{(b)} and \ref{fig:NV_triplets_PBE_vs_DDH}\textbf{(c)}, respectively.
 
\begin{figure}[!h]
    \begin{minipage}[c]{1.0\linewidth}
    \centering
    \includegraphics[width=1.0\textwidth]{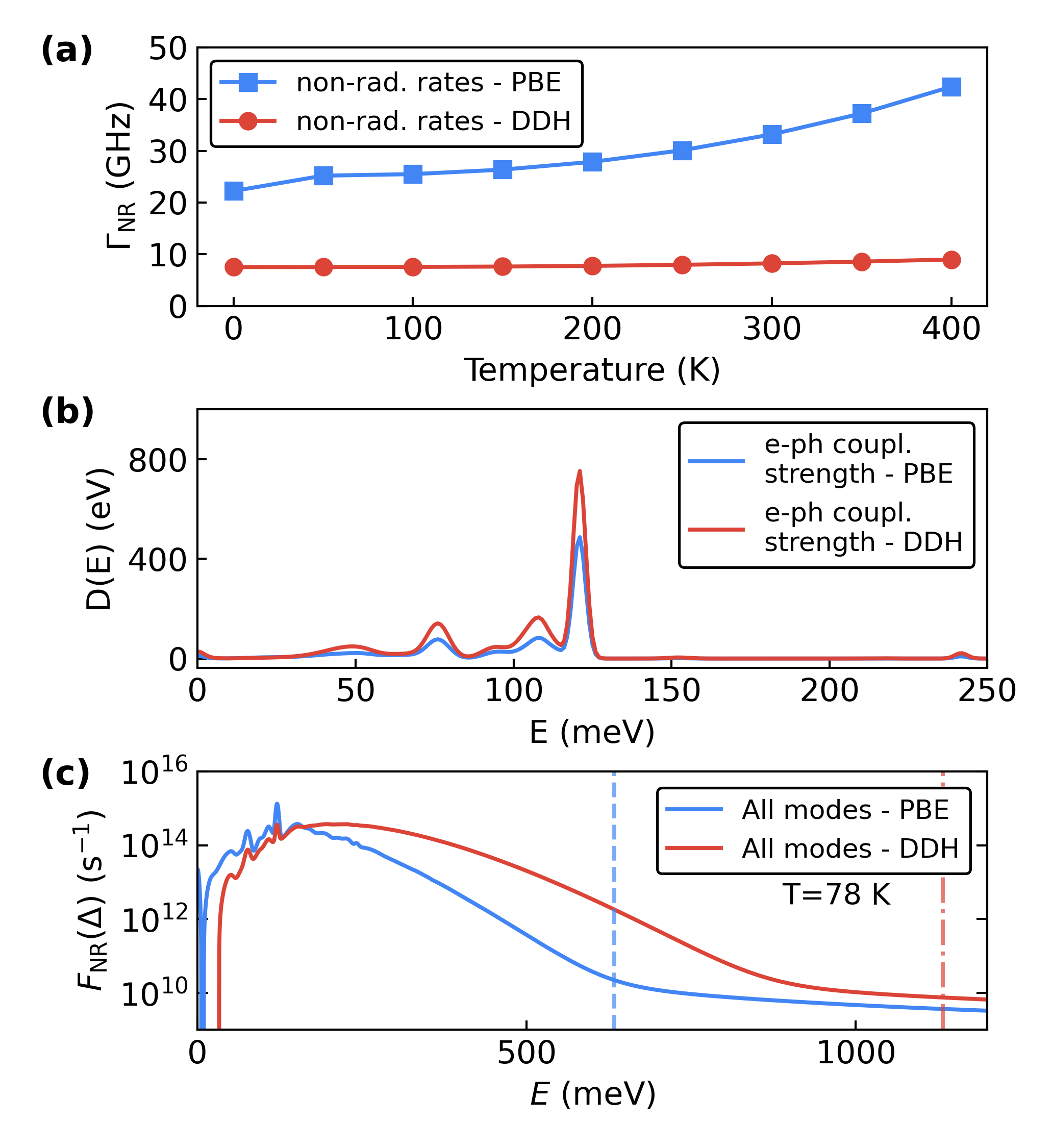}
    \end{minipage}
    \caption{\textbf{(a)}: Non-radiative rates of the ${}^1\mathrm{A}_1\to{}^1\mathrm{E}$ transition of the \textit{kk}-VV$^0$ in 4H-SiC for temperatures in the range $T=0\sim300~\mathrm{K}$ obtained using the PBE and DDH functionals. \textbf{(b)}: Comparison between the e-ph coupling spectral density functions. \textbf{(c)}: Comparison between the line-shape functions. The blue dashed line and the red dotted-dashed line indicate the adiabatic energy differences computed with TDDFT and the PBE and DDH functionals, respectively. To date, there are no experimental estimates of the ZPL.}
    \label{fig:SiC_singlets_PBE_vs_DDH}
\end{figure}

Next, we consider the ${}^1\mathrm{A}_1\to{}^1\mathrm{E}$ transition of the \textit{kk}-VV$^0$ in 4H-SiC. We computed the NRTRs with PBE and DDH using the corresponding adiabatic energy differences $\Delta_{IF}=0.633~\mathrm{eV}$ and $\Delta_{IF}=1.132~\mathrm{eV}$ obtained from TDDFT\cite{jin2023excited}, and present our results in Figure \ref{fig:SiC_singlets_PBE_vs_DDH}. We observe noteworthy differences between the results obtained with the two functionals; differences are also found in the e-ph spectral densities and the line-shape functions, shown in Figures \ref{fig:SiC_singlets_PBE_vs_DDH}\textbf{(b)} and \ref{fig:SiC_singlets_PBE_vs_DDH}\textbf{(c)}, respectively. We attribute this difference to the different adiabatic energies predicted by the two functionals.
% \begin{table}[ht]
% \centering
% \caption{Comparison of non-radiative transition rates between the singlet states of the \textit{kk}-VV${^0}$ center in 4H-SiC obtained from first-principles with PBE and DDH functionals. $\Delta_{IF}$ is the adiabatic energy difference.}
% \begin{tabular}{ccccc}
% \toprule
% \toprule
% % \hline \hline
% & & \multicolumn{3}{c}{$\Gamma_{\mathrm{NR}}^{{}^1\mathrm{A}_1,{}^1\mathrm{E}}$  (GHz)} \\
% \addlinespace[2pt]
% & $\Delta_{IF}$ (eV) & $T = 4\,\mathrm{K}$ & $T = 78\,\mathrm{K}$ & $T = 300\,\mathrm{K}$\\
% PBE & 0.633 & $25.19$ & $25.30$ & $33.19$ \\
% DDH & 1.132 & $7.53$ & $7.54$ & $8.24$ \\
% \bottomrule
% \bottomrule
% \end{tabular}
% \label{tab:NRTR_singlets_SiC}
% \end{table}
%

\begin{figure}[!htb]
    \begin{minipage}[c]{1.0\linewidth}
    \centering
    \includegraphics[width=1.0\textwidth]{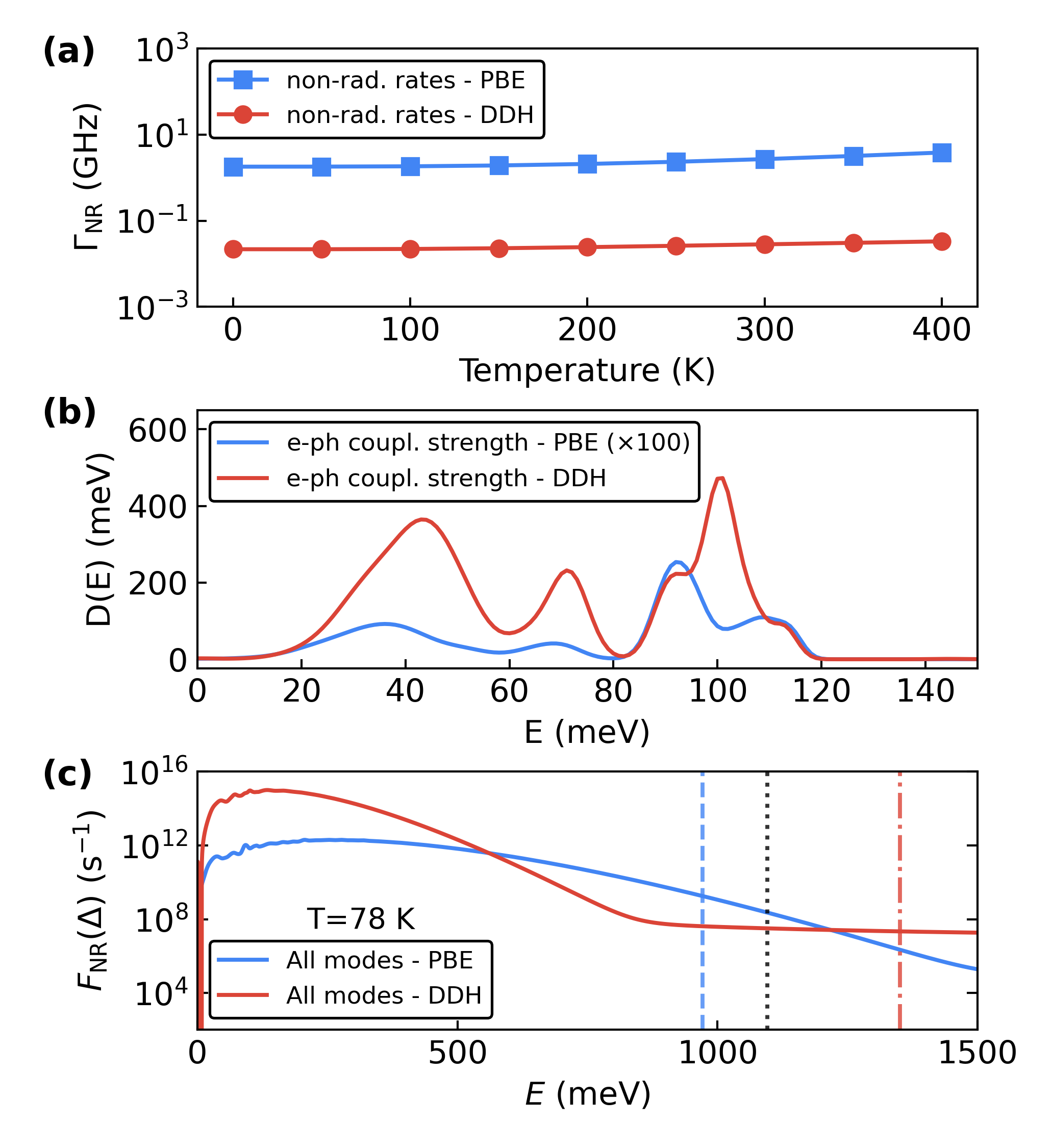}
    \end{minipage}
    \caption{\textbf{(a)}: Non-radiative rates of the ${}^3\mathrm{E}\to{}^3\mathrm{A}_2$ transition of the \textit{kk}-VV$^0$ in 4H-SiC for temperatures in the range $T=0\sim400~\mathrm{K}$ obtained using the PBE and DDH functionals. \textbf{(b)}: Comparison between the e-ph coupling spectral density functions. \textbf{(c)}: Comparison between the line-shape functions. The blue dashed line and the red dotted-dashed line indicate the adiabatic energy differences computed with TDDFT and the PBE and DDH functionals, respectively. The black dotted line indicates the experimental value of the ZPL.}
    \label{fig:SiC_triplets_PBE_vs_DDH}
\end{figure}

\begin{figure}[!h]
    \begin{minipage}[c]{1.0\linewidth}
    \centering
    \includegraphics[width=1.0\textwidth]{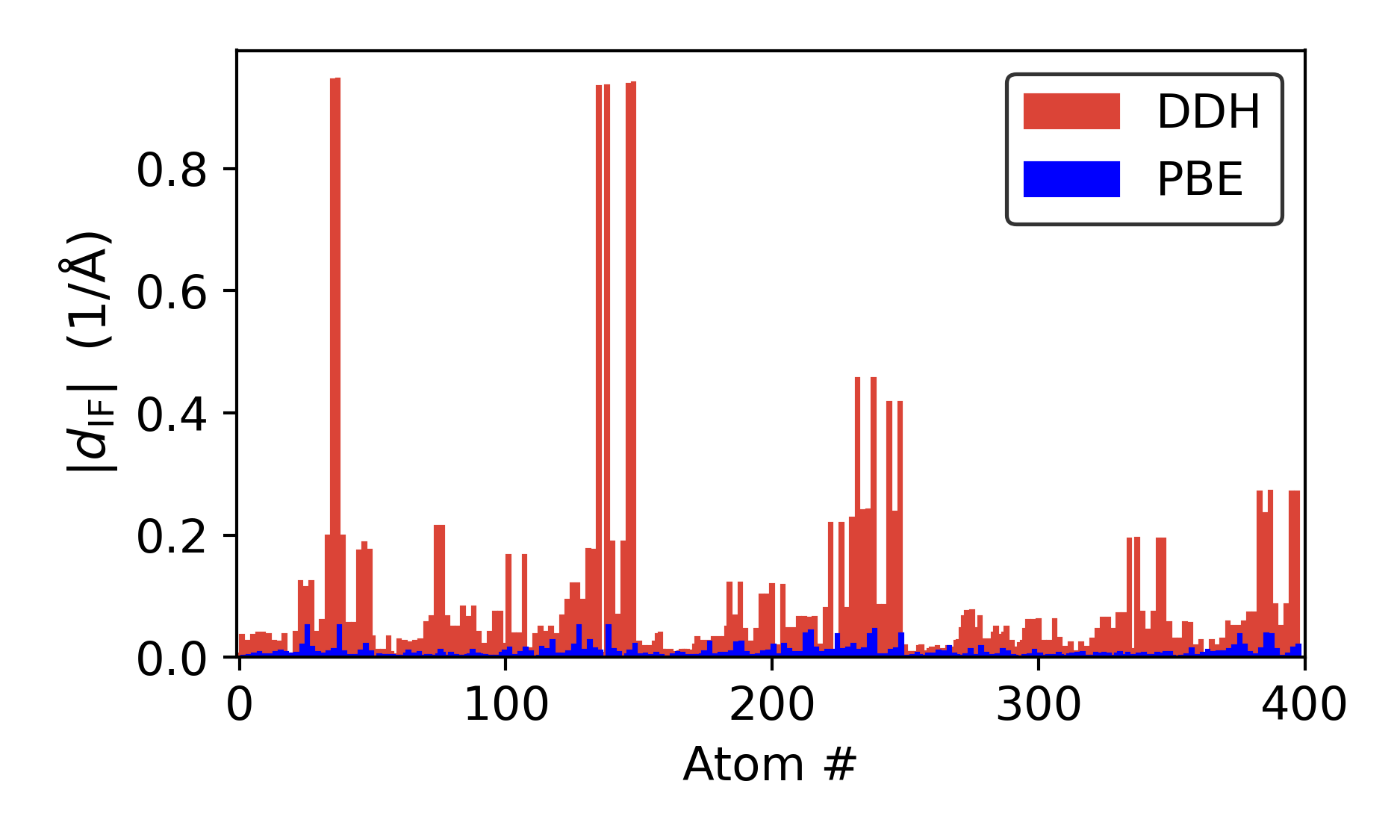}
    \end{minipage}
    \caption{Absolute values of the NAC vectors $\bs{d}_{IF,A}$ between the ${}^3\mathrm{E}$ and ${}^3\mathrm{A}_2$ states of the \textit{kk}-VV$^0$ center in 4H-SiC computed for each of the 398 atoms in the simulation cell using the DDH and PBE functionals on the respective excited-state relaxed geometries.}
    \label{fig:SiC_NACs_PBE_vs_DDH}
\end{figure}

\begin{figure}[!h]
    \begin{minipage}[c]{1.0\linewidth}
    \centering
    \includegraphics[width=1.0\textwidth]{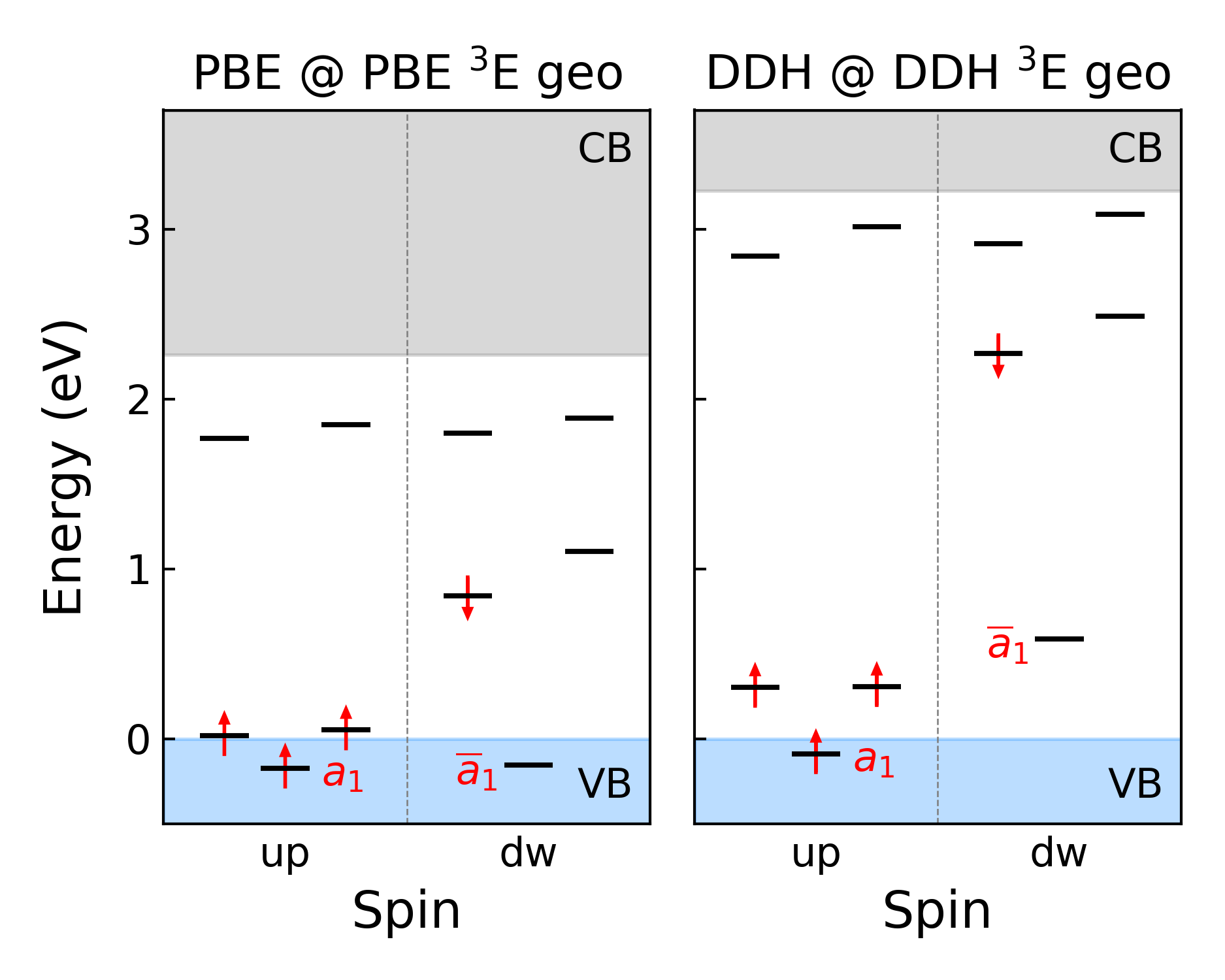}
    \end{minipage}
    \caption{Position of the single-particle defect levels for the \textit{kk}-VV$^0$ center in 4H-SiC computed using the PBE and the DDH functionals and the relaxed atomic geometries obtained, respectively, minimizing the PBE and DDH TDDFT total energy of the ${}^3\mathrm{E}$ state.}
    \label{fig:SiC_orbitals_energy_PBE_vs_DDH}
\end{figure}

Finally, we present results for the ${}^3\mathrm{E}\to{}^3\mathrm{A}_2$ transition of the \textit{kk}-VV$^0$ in 4H-SiC, where we find large differences in the NRTRs computed with the two functionals, as displayed in Figure \ref{fig:SiC_triplets_PBE_vs_DDH}\textbf{(a)}. The strong dependence on the functional is also evident in the e-ph spectral densities, in Figure \ref{fig:SiC_triplets_PBE_vs_DDH}\textbf{(b)}, and in the line-shape functions, in Figure \ref{fig:SiC_triplets_PBE_vs_DDH}\textbf{(c)}. The difference between $\Delta_{IF}$ values ($0.971$ eV and $1.371$ eV with PBE and DDH, respectively) is not the only factor responsible for the differences in results found with the two functionals.  The two e-ph spectral densities $D(E)$ show quantitative differences not only in the relative heights of the peaks, but also in the magnitude of the coupling strength. A detailed investigation reveals that this discrepancy arises from the NACs, whose values differ substantially between the two functionals, as shown in Figure \ref{fig:SiC_NACs_PBE_vs_DDH}. We find that at the PBE level, inaccuracies in the description of the ${}^3\mathrm{E}$ state and its relaxed geometry lead to a NAC remarkably different from that obtained with DDH. In Figure \ref{fig:SiC_orbitals_energy_PBE_vs_DDH} we compare the defect's single-particle levels computed using PBE and DDH at the respective relaxed geometrical configurations. We find that PBE predicts the energy of the $\overline{a}_1$ defect orbital to be below the valence-band maximum, in contrast with the expected behavior captured by the DDH functional, which predicts the orbital energy to be in the band gap of the host. Furthermore, we computed the orbital localization factors using the WEST code and found that the $\overline{a}_1$ orbital obtained using PBE and the PBE relaxed geometry has highly delocalized character, in contrast with the expected behavior, which is found in the localized character of the orbital obtained with DDH at the DDH relaxed geometry. We argue that this incorrect description of the ${}^3\mathrm{E}$ state character and of the relaxed geometry stems from known inaccuracies of the PBE functional in describing charge-transfer processes in spin-defective systems.

\section{Computational details and workflow description}
All the DFT calculations were carried out with the Quantum ESPRESSO code\cite{giannozzi2017advanced}. For both systems, we used a kinetic energy cutoff of 60 Ry for the wavefunctions and optimized norm-conserving Vanderbilt (ONCV) pseudo-potentials\cite{schlipf2015optimization}. All TDDFT calculations\cite{jin2023excited} were performed with the WEST code, which implements analytical formulations of excited-state forces\cite{jin2023excited} and NACs\cite{villani2026}. We generated the dynamical matrices using the phonopy code\cite{togo2015first}. To reduce the computational cost required to obtain excited-state phonons with finite displacements in large supercells, we computed all the dynamical matrices at the PBE level of theory, following a practice justified in previous works\cite{jin2021photoluminescence,kundu2024quantum}. We assumed no mode mixing between the vibrational modes of the initial and final electronic states and we employed the displaced harmonic oscillator approximation throughout. Specifically, we considered the excited-state phonons of the initial excited state for all the transitions analyzed except for the ${}^3\mathrm{E}\to{}^3\mathrm{A}_2$ transition in SiC, where instead we used the ground-state phonons. We note that the relaxed atomic configuration of the ${}^3\mathrm{E}$ state in \textit{kk}-VV$^0$ in 4H-SiC possesses a broken-symmetry geometry with $C_{1h}$ point group and computing the dynamical matrix of the excited state in the 398-atom supercell requires thousands of TDDFT-forces calculations, a prohibitively expensive task even with the PBE functional. Line-shape functions were computed using the PyPL code\footnote{\url{https://github.com/jinyuchem/pypl}}. For each transition, we used values of the Lorentzian smearing factor $\gamma$ appearing in the generating function $G(E)$ estimated from low-temperature photoluminescence linewidth measurements. For the ${}^1\mathrm{A}_1\to{}^1\mathrm{E}$ transition in NV$^-$ we used $\gamma=0.18~\mathrm{meV}$\cite{biktagirov2017strain}; for the ${}^3\mathrm{E}\to{}^3\mathrm{A}_2$ transition in NV$^-$ we used $\gamma=2.07\times 10^{-4}~\mathrm{meV}$\cite{fu2009observation}; for the ${}^3\mathrm{E}\to{}^3\mathrm{A}_2$ transition in \textit{kk}-VV$^0$ in 4H-SiC we used $\gamma=4.13\times10^{-4}~\mathrm{meV}$\cite{christle2017isolated}; for the ${}^1\mathrm{A}_1\to{}^1\mathrm{E}$ transition in \textit{kk}-VV$^0$ in 4H-SiC, where no experimental data are available, we used the same $\gamma=0.18~\mathrm{meV}$ used for the NV$^-$.

\clearpage
\bibliographystyle{apsrev4-2}
\bibliography{bibliography}